# Dark Patterns in the Opt-Out Process and Compliance with the California Consumer Privacy Act (CCPA)

VAN TRAN, University of Chicago, USA
AARUSHI MEHROTRA, University of Chicago, USA
RANYA SHARMA, University of Chicago, USA
MARSHINI CHETTY, University of Chicago, USA
NICK FEAMSTER, University of Chicago, USA
JENS FRANKENREITER, Washington University in St. Louis, USA
LIOR STRAHILEVITZ, University of Chicago, USA

To protect consumer privacy, the California Consumer Privacy Act (CCPA) mandates that businesses provide consumers with a straightforward way to opt out of the sale and sharing of their personal information. However, the control that businesses enjoy over the opt-out process allows them to impose hurdles on consumers aiming to opt out, including by employing dark patterns. Motivated by the enactment of the California Privacy Rights Act (CPRA), which strengthens the CCPA and explicitly forbids certain dark patterns in the opt-out process, we investigate how dark patterns are used in opt-out processes and assess their compliance with CCPA regulations. Our research reveals that websites employ a variety of dark patterns. Some of these patterns are explicitly prohibited under the CCPA; others evidently take advantage of legal loopholes. Despite the initial efforts to restrict dark patterns by policymakers, there is more work to be done.

## 1 Introduction

The California Consumer Privacy Act (CCPA), enacted in 2018, is a significant legislative stride in US consumer privacy protection law [22]. The right to opt out of the sale and sharing of personal information is a cornerstone provision within the CCPA, allowing consumers to prevent businesses from selling or sharing their personal information. In practice, however, the effectiveness of this right might be undermined when businesses implement dark patterns in the opt-out process. Dark patterns are manipulative user interfaces designed to trick, deceive, or confuse users into taking actions that benefit the business rather than the consumer. These tactics can make it difficult for consumers

Authors' Contact Information: Van Tran, University of Chicago, Chicago, IL, USA, tranv@uchicago.edu; Aarushi Mehrotra, University of Chicago, Chicago, IL, USA, aarushi.mehrotra0@gmail.com; Ranya Sharma, University of Chicago, Chicago, IL, USA, ranyasharma@uchicago.edu; Marshini Chetty, University of Chicago, Chicago, IL, USA, marshini@uchicago.edu; Nick Feamster, University of Chicago, Chicago, IL, USA, feamster@uchicago.edu; Jens Frankenreiter, Washington University in St. Louis, St. Louis, MO, USA, fjens@wustl.edu; Lior Strahilevitz, University of Chicago, Chicago, IL, USA, lior@uchicago.edu.

2 Yan Tran, Aarushi Mehrotra, Ranya Sharma, Marshini Chetty, Nick Feamster, Jens Frankenreiter, and Lior Strahilevitz

to successfully exercise their opt-out rights [3, 35] (we are going to refer to the process of exercising these rights as the opt-out process in the rest of the paper). To address these limitations, the California Privacy Rights Act (CPRA) introduced additional clarifications on opt-out requirements and explicitly prohibited the use of dark patterns in the opt-out process. These provisions represent the first American law to explicitly ban various dark patterns [6]. However, questions remain about the law's effectiveness in preventing websites from using these tactics. Since the CCPA serves as the overarching statutory framework encompassing both the CCPA and CPRA, we will use the term "CCPA" to refer to the comprehensive set of privacy regulations, including opt-out requirements.

To implement the opt-out right, businesses that are subject to the CCPA that collect personal information from consumers online shall, *"at minimum, allow consumers to submit requests to opt-out of sale/sharing through an opt-out preference signal and at least one of the following methods: an interactive form accessible via the 'Do Not Sell or Share My Personal Information' link, the Alternative Opt-out link, or the business's privacy policy if the business processes an opt-out preference signal in a frictionless manner"* [4][1]. In this study, we use the term "Opt-out link" to refer to both the "Do Not Sell or Share My Personal Information" link and the Alternative Opt-out link. The opt-out link requires consumers to manually click a designated link on a website's homepage and submit an individual opt-out request for each site. In contrast, frictionless opt-out preference signals, once set up, are automatically sent from the user's device or browser to websites, allowing for seamless opt-out. Previous studies have noted that the opt-out process via the opt-out link is often burdensome [3] and that websites more frequently implement the opt-out link than frictionless opt-out preference signals [11, 41]. Therefore, this study focuses on the opt-out process using the opt-out link.

Building on previous research that examined the CCPA opt-out process before the enactment of the CPRA [3, 10, 21, 34], we conduct one of the first studies concerning the opt-out process of CCPA-subject websites after the CPRA went into effect. More importantly, we analyze the dark patterns present in the opt-out process and compare them with the CCPA regulations to determine whether these dark patterns violate the CCPA regulations or whether they exploit the legal gaps in the CCPA. Here are the research questions addressed in the study, along with a summary of the results:

- *RQ1: Can consumers successfully exercise their opt-out rights? (Section 4.2)* Nearly 30% of the attempted opt-out requests could not be submitted, did not succeed, or had unclear status updates.
- *RQ2: What does it take for a consumer to submit an opt-out request? (Section 4.3) Do opt-out processes violate CCPA requirements? (Section 4.4)* The opt-out process varies widely depending on website categories, opt-out control methods, and third-party platforms handling the opt-out process on behalf of the websites. Some sites allow submission with just a few clicks, while others require a lengthy process and significant personal information.
- *RQ3: What types of dark patterns are used in the opt-out process? (Section 5.2) Are these dark patterns violating CCPA requirements or are they exploiting the loopholes that have not been addressed by the CCPA? (Section 5.3)* Multiple types of dark patterns are found in the opt-out process, which can be grouped into three categories: "Obstruction", "Interface interference", and "Misdirection". Although some dark patterns are clearly prohibited by the CCPA, others are not explicitly addressed by the CCPA.
- *RQ4: What are the characteristics of the websites that employ dark patterns? (Section 5.4)* Most websites implement at least one type of dark pattern. Sites with different opt-out control methods and third-party platforms tend to exhibit different types of dark patterns.

To answer these questions, we systematically exercised opt-out rights and documented the opt-out process for 330 websites subject to the CCPA, across various categories such as "Retail", "Finance", "Apparel", and others. We tracked the outcome of the opt-out requests and examined the types of dark patterns present in the process. In summary, we make the following contributions:

(1) *We study website behaviors post-CPRA.* Hence, it offers timely insights into how websites are adapting their opt-out processes to comply with this privacy law.
(2) *We systematically exercise, record, and characterize the CCPA opt-out process.* Unlike previous studies that focus only on certain parts of the opt-out process, such as the opt-out page, we document the full process, including all actions required after the request is submitted. This comprehensive approach enables us to accurately and thoroughly identify implementation patterns, requirements, and any dark patterns present in the opt-out process.
(3) *We study dark patterns found in the CCPA opt-out process and assess whether they violate explicit prohibitions in the CCPA.* Because our study focuses on CCPA-subject websites after the CPRA took effect, our findings on dark patterns offer important insights into potential compliance or non-compliance with the CCPA. We conducted a detailed comparison between our observations and the CCPA regulations to determine where websites align with or deviate from legal standards. Furthermore, our analysis highlights potential loopholes in the CCPA that some businesses may exploit to hinder consumers from effectively exercising their opt-out rights.
(4) *We provide a robust, well-labeled dataset that offers valuable insights into the CCPA opt-out process and its associated dark patterns.* We developed a comprehensive, well-organized dataset on opt-out processes and dark patterns, offering up-to-date insights into how these processes are implemented. This dataset is a valuable resource for researchers, regulators and policymakers, shedding light on current practices, dark patterns, and potential weaknesses in CCPA regulations.

---

[1] *"Italicized text within quotations"* indicates excerpts from CCPA provisions.

The rest of the paper is organized as follows. We start by providing a review of related work in Section 2, then we explain our dataset selection and methodology in Section 3. Our findings consist of two parts. We present findings on opt-out request outcomes and the opt-out process in Section 4 and findings about dark patterns in Section 5. We discuss the implications of our results in Section 6 and conclude in Section 7.

## 2 Related Work

This section presents related work about dark patterns, legal compliance, and CCPA opt-out rights.

### 2.1 Dark Patterns in User Interface Design

Since the term "dark patterns" was formally introduced in 2010, there has been a growing body of scholarship examining these deceptive designs across various domains and contexts. Researchers have shown that dark patterns are prevalent on numerous platforms, devices, and website types, including e-commerce sites, game designs, and IoT devices [27, 30, 31]. Dark patterns can serve multiple purposes. Some can trick consumers into purchasing unwanted products or make it difficult to discontinue services or delete accounts [10, 17]. Others can harm consumers by compromising their privacy. Dark patterns may include designs that deceive users into revealing excessive personal information or complicate the process of opting out of cookie tracking [9, 28, 29, 32, 40]. A major concern with dark patterns is that they can have a significant impact on changing consumers' behaviors [15, 27]. Our study contributes to this literature by examining dark patterns in the CCPA opt-out process.

4 Van Tran, Aarushi Mehrotra, Ranya Sharma, Marshini Chetty, Nick Feamster, Jens Frankenreiter, and Lior Strahilevitz

## 2.2 Compliance With Privacy Laws

Given the rapid increase in the collection, sale, and distribution of personal data, there has been growing interest in studying consumer privacy. Various privacy laws have been enacted, including sector specific privacy laws like the Health Insurance Portability and Accountability Act (HIPAA), Gramm-Leach-Bliley Act (GLBA), and Children's Online Privacy Protection Act (COPPA), as well as federal and state-level privacy laws such as the CCPA in the US and the General Data Protection Regulation (GDPR) in the European Union. Yet, various studies looking at apps' and websites' implementation of opt-out requirements, including privacy controls, privacy notices, and privacy policies [12, 33, 34, 36, 37, 43], all point to the same conclusion: Many businesses either fail to implement these requirements or use tactics that confuse or deceive consumers, making it challenging for them to exercise their privacy rights [14, 33, 38, 42]. For example, Sanchez et al. [38] found that although the GDPR reduces the amount of tracking by websites, many websites may present users with deceptive information that makes avoiding tracking difficult. Nouwens et al. [33] also found that dark patterns and implied consent are prevalent in cookie consent interfaces and consent management platforms. These findings raise concerns about the true effectiveness of these privacy laws, since businesses can implement tactics that discourage consumers from exercising their privacy choices. Our study examines the CCPA opt-out process on CCPA-subject websites following the implementation of the CPRA. Since the CPRA imposes various requirements for the opt-out process as well as explicitly prohibits the use of certain dark patterns, our findings provide insights into website compliance with CCPA requirements. The evidence gathered can guide regulators in enforcement actions and help policymakers strengthen regulations to prevent businesses from exploiting potential loopholes.

## 2.3 Dark Patterns in CCPA Opt-out Processes

Among various privacy laws, the CCPA, as the first comprehensive privacy law in the United States [6], has garnered significant attention from policymakers and the CHI community. It not only enforces consumer data rights in California but also serves as a model for other states, setting a new national standard for privacy protection and shaping the future of data regulation across the country [7, 8]. Another distinction that sets the CCPA (specifically the CPRA) apart is that it is the first U.S. law to address concerns about dark patterns [6]. Consequently, it plays a crucial role in shaping user interfaces to promote transparency and prevent manipulative practices that undermine consumer privacy rights.

Researchers have explored various aspects of the CCPA. Some have focused on measuring the presence of opt-out methods, such as opt-out links or preference signals [34, 43, 44], while others have examined the design choices in the opt-out process and their impact on consumers [3, 10, 12, 19, 21, 23, 39]. For example, studies have looked at how the language and icons used to present privacy options [13, 21], as well as the visibility of opt-out choices [39], influence consumers' understanding and decisions. Other research has analyzed how different designs for opt-out controls—such as offering a single option, multiple options, or a fillable form—affect consumer behavior [35]. A common finding across these studies is that design choices significantly impact both consumers' comprehension and their willingness to opt out [18, 20, 24, 35].

A few studies have examined the implementation of the CCPA opt-out process and how the opt-out process allows users to exercise their rights, mostly before the CPRA went into effect. For example, in 2020, Consumer Reports conducted a study in which over 500 California residents were asked to exercise their CCPA opt-out rights and share their experiences. They found that "many data brokers' opt-out processes are so onerous that they have substantially impaired consumers' ability to opt out, highlighting serious flaws in the CCPA's opt-out model", and "52% of the time, the tester was 'somewhat dissatisfied' or 'very dissatisfied' with the opt-out processes" [3]. Additionally, some research [34] has attempted to study CCPA dark patterns. Unfortunately, this work focuses solely on the limited types of dark patterns found on the opt-out page, without considering the entire opt-out process, which is often more complex and may involve a wider range of dark patterns. In this study, we systematically record and engage with the entire opt-out process, allowing us to accurately document the full extent of dark patterns, not just on the opt-out page but throughout the entire process. The evidence collected provides valuable insights into CCPA-related dark patterns, serving as a valuable resource for researchers, regulators, and policymakers.

## 3 Method

In this section, we first outline our method for selecting the websites to analyze. Then, we describe the empirical design, which has three parts:

(1) *Develop metrics & dark patterns taxonomy:* First, in January 2023, we conducted a pilot study on 100 websites randomly selected from the set of websites to analyze. The objective of the pilot study was to develop metrics for characterizing the opt-out process, create a taxonomy to describe this process, and design an opt-out questionnaire for the actual data collection.
(2) *Data collection & labeling:* Next, from March 12, 2024, to March 30, 2024, we conducted the main data collection and labeling process.
(3) *Follow-up with request:* Finally, we completed any additional steps required by some websites to ensure that opt-out requests were successfully executed.

### 3.1 Website Selection

The process of selecting the set of popular websites that are subject to the CCPA and contain the opt-out link proceeds as follows:

(1) We first used stratified sampling to obtain the top 1,000 websites across various categories—such as "Finance", "Retail", "Apparel", and "Automotive"—that have at least 1 million website visitors. This data (including categories of websites) was sourced from a reputable website-ranking platform [25].
(2) In January 2024, using an IP address based in California, we manually checked each website's homepage for the presence of an opt-out link. We identified 542 websites that met this criterion, and websites without an opt-out link were excluded from the dataset.
(3) Next, we filtered the websites to identify those likely subject to the CCPA, following the methodology outlined in Tran et al. [41]. We selected only for-profit businesses with annual gross revenues of at least $25 million and have U.S. branch locations (and therefore, likely to have California consumers), resulting in a final dataset of 330 websites.

### 3.2 Opt-out Metrics & Dark Patterns Taxonomy Development

*3.2.1 Characterization metrics.* The implementation of the opt-out process varies significantly across websites. Some sites allow users to submit a request with just one or a few clicks, while others require a multi-step form and the submission of extensive personal information. To capture this variability, we developed three specific metrics to characterize what it takes for a consumer to opt-out.

Van Tran, Aarushi Mehrotra, Ranya Sharma, Marshini Chetty, Nick Feamster, Jens Frankenreiter, and Lior Strahilevitz

**1. Opt-out control methods:** Opt-out control methods for consumers can generally be categorized into three types which we detail below.

- *Browser opt-out:* After consumers click on the opt-out link, they can select certain opt-out options—such as toggling off data sharing or the sale of personal information—without needing to provide any personal details. We call this control method "browser opt-out" since it is likely to use a browser-based method (not personal details) to identify users.
- *Opt-out form:* Consumers must fill out an opt-out form that requires them to provide personal information, such as their name, email address, phone number, or mailing address. This process typically involves multiple steps, including submitting the form and possibly confirming the request via email or other means.
- *Both browser opt-out & opt-out form (Both control methods):* In this scenario, consumers must use browser opt-out in addition to completing an opt-out form where they have to provide personal information.

**2. Number of clicks:** We count the number of clicks as a proxy for how many steps are required to submit an opt-out request. To ensure counting of the number of clicks is consistent, we only count the necessary clicks that must be performed to submit a request, not including those used to solve a CAPTCHA, those used to move between different fields in the opt-out form, or those used to explore other sections not related to the opt-out process.

**3. Personal information required:** We note whether the following four items of personal information are required to complete the opt-out process: name, email address, phone number, and physical address.

*3.2.2 Dark Patterns Taxonomy.* We draw from two key sources in our approach: (1) existing dark pattern taxonomies developed by the Federal Trade Commission (FTC) [26], the European Data Protection Board [1], and research like that of Gray et al. [16], and (2) findings from our pilot study on the opt-out process. Using these resources, we classified potential dark patterns in the opt-out process into three main groups, which we refer to as high-level patterns: "Obstruction", "Interface interference", and "Misdirection". We refined these categories further by introducing low-level patterns, which are shorthand terms that describe the specific traits of a dark pattern. For example, "Privacy maze" ("Obstruction" dark pattern) refers to a complex opt-out process that requires navigating through multiple steps. In cases where no existing terms were adequate, we developed new shorthand terms to clearly define the dark patterns we observed.

Our taxonomy extends beyond the dark patterns explicitly mentioned in CCPA regulations. This broader framework allows us not only to assess compliance with CCPA requirements but also to identify dark patterns that are not directly addressed by the regulations. For the full taxonomy of dark patterns used in this paper, refer to Section 5.2.

*3.2.3 Opt-out Questionnaire.* The questionnaire primarily consists of multiple-choice questions designed to capture participants' observations regarding the opt-out process and any dark patterns encountered. It is divided into two sections. Part 1 addresses general aspects of the opt-out process, including questions about opt-out control methods, personal information requested, and whether third-party platforms are used to handle opt-out process. Part 2 focuses specifically on identifying any dark patterns present during the opt-out process, with participants indicating whether or not they observed each type of dark pattern.

## 3.3 Data Collection & Labelling

Since the CCPA applies only to Californians, we conducted the study from a California vantage point. Due to our lack of access to a physical machine in California, we used a Virtual Private Network (VPN) to simulate browsing from within the state. Some websites require personal information for opt-out requests, so we created a California resident profile with a fake name, email address, California phone number, and a publicly available California address. We did not create accounts nor login for any websites since the CCPA prohibits businesses from requiring a consumer to create an account to opt out [4].

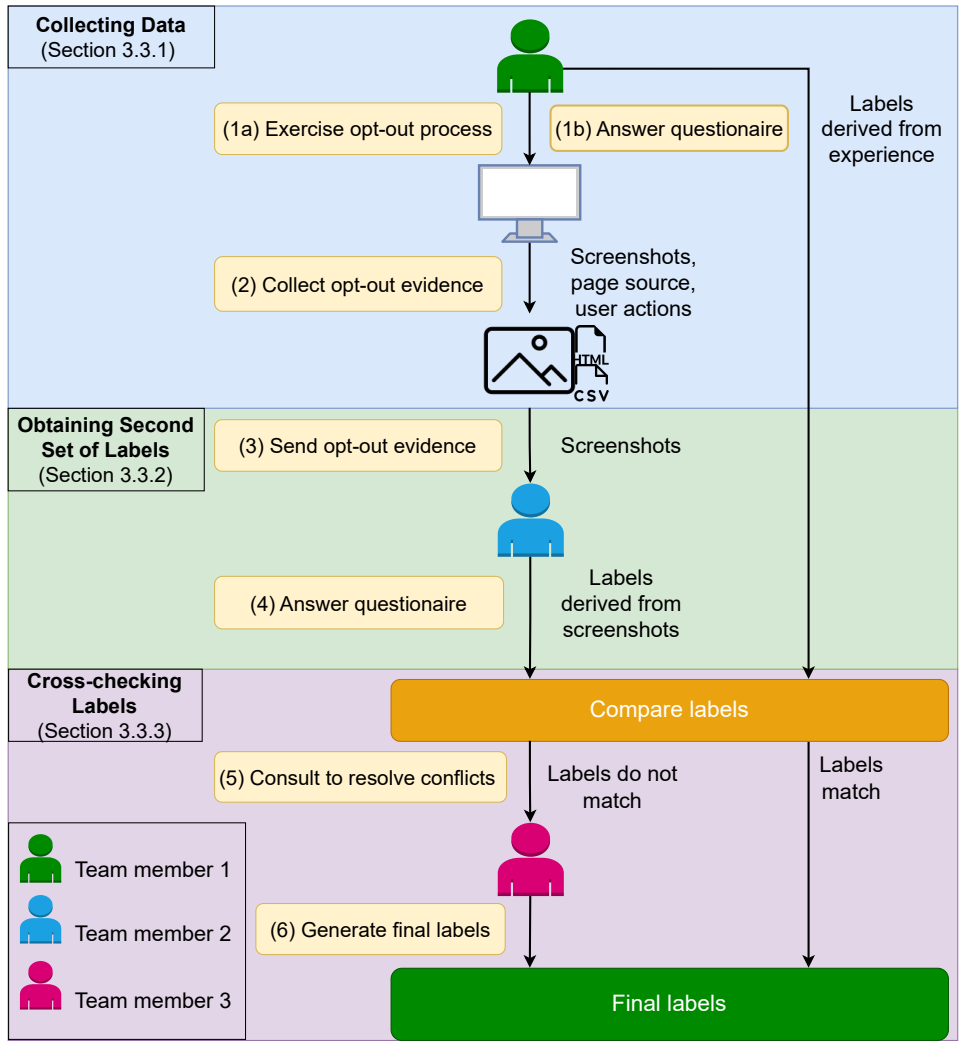

Fig. 1. Pipeline to collect and label opt-out evidence.

The data collection and labeling process involved three team members. Team Member 1 collected the data and labeled the opt-out process and dark patterns based on her observations while completing the opt-out process. Team Member 2 8 Van Tran, Aarushi Mehrotra, Ranya Sharma, Marshini Chetty, Nick Feamster, Jens Frankenreiter, and Lior Strahilevitz counted the number of clicks taken to submit a request[2] and labeled the opt-out process and dark patterns based on the screenshots provided by Team Member 1. Team Member 3 reviewed any discrepancies between the labels to finalize the labels. Figure 1 illustrates the data collection and labeling process, with each "step" referenced below corresponding to those in the figure.

*3.3.1 Collecting Data.* Team Member 1 exercised the opt-out process on all of 330 websites using our custom data collection pipeline (Step 1a).[3] During the opt-out process, Team Member 1 labeled the opt-out process and dark patterns encountered during the process using the opt-out questionnaire (Step 1b). This provided the first set of labels for the dark patterns encountered. The data collection pipeline, built on the Python library Selenium, automatically records every mouse click and keyboard action performed and save them in a CSV file at the end of the opt-out process. Additionally, each mouse click automatically triggers screenshots and page source captures. In summary, opt-out evidence collected (Step 2) by the pipeline includes:

(1) A CSV file with all recorded mouse and keyboard actions.
(2) Screenshots taken at every click.
(3) The page source captured at each click.

*3.3.2 Obtaining Second Set of Labels.* Screenshots collected from the pipeline were then sent to Team Member 2 (Step 3) who answered the questionnaire (Step 4) but without referencing Team Member 1's responses. This provided a second independent set of labels for the dark patterns encountered. Additionally, Team Member 2 counted the number of clicks required to complete the opt-out process.

*3.3.3 Cross-checking Labels.* We compared the dark patterns labels by Team Member 1 & 2. If there was no disagreement, those labels were finalized for the website. If there were any discrepancies, we asked Team Member 3 to review the case (Step 5) and finalize the labels (Step 6). We used Cohen's $\kappa$ score to assess the level of agreement between labels by Team Member 1 & 2. We found that half of the dark patterns cross-checked have substantial or almost perfect agreement with each other ($\kappa >= 0.6$). However, some "Interface interference" and "Misdirection" dark patterns only have slight agreement or moderate agreement with each other ($0 <= \kappa < 0.6$). We hypothesize that certain dark patterns show higher agreement between labels due to their objective nature—focusing on whether specific actions or features are required, present, or absent in the opt-out process—while patterns with lower agreement may be more subjective. Refer to Table 10 in the Appendix for more details.

## 3.4 Request Follow-up Process

Some websites require users to complete verification steps after submitting an opt-out request. To track this, Team Member 1 collected all emails exchanged with the website for two months following the submission. (Note: The CCPA requires businesses to respond within 15 days.)[4] For each email received, Team Member 1 summarized the required actions and the email's purpose, recording this in an Excel file. At the end of the two-month period (June 2024), Team Member 2 reviewed the Excel file, coded the actions taken, and analyzed the final email from the website to determine the outcome of the opt-out request.

---

[2]We initially attempted to automatically count the number of clicks, but this proved unreliable. Consumers often click to explore other sections or navigate between fields in the opt-out form, leading to inconsistent results.
[3]If she failed to complete the opt-out process for any of the websites, Team Member 2 would attempt to complete the opt-out process for that website a few days later (not more than 7 days) to verify the existence of these issues with completing the opt-out process for that website.
[4]Each website was assigned a unique identifier, and email sub-addressing was used to link this identifier to our communications.

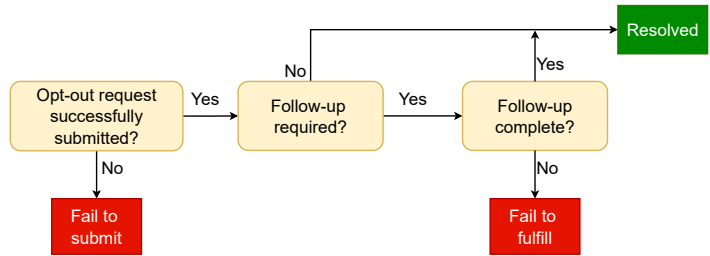

Fig. 2. Categorizing opt-out request outcomes.

## 4 Findings About Opt-out Process

This section presents our findings on the CCPA opt-out process. We first start with an overview of the CCPA's requirements for the opt-out process. After that, we present findings related to the opt-out process including: (1) whether consumers can successfully exercise their opt-out rights, and (2) what is required for a consumer to submit an opt-out request. For each finding, we also describe the CCPA's requirements and identify potential legal violations.

### 4.1 What Requirements Does the CCPA Impose on the Opt-Out Process?

The CCPA sets specific requirements for the opt-out process. Key requirements include:

- *Opt-out process:* The CCPA explicitly requires that the opt-out process uses an *"online interactive form"* that is *"easy for consumers to execute and require minimal steps"* [4].
- *Verification of request:* *"Businesses shall not require a consumer to verify their identity to make a request to opt out"* [4]. However, businesses may ask consumers for personal information if it is *"necessary to direct the business not to sell or share the consumer's personal information"* [4].
- *Denial of request:* Although businesses may deny the request if they have a *"good-faith, reasonable and documented"* belief that the request is fraudulent, they *"shall inform the requester that it will not comply with the request and provide an explanation to the requester why it believes the request is fraudulent"* [4].
- *Response timeline:* Businesses have to comply with the request by ceasing the sale/sharing of personal information within 15 business days from the date they receive the request [4].

In addition to mandatory requirements, businesses may choose to implement optional provisions, such as providing a way for consumers to confirm that their request to opt-out of sale/sharing has been processed: *"A business may provide a means by which the consumer can confirm that their request to opt-out of sale/sharing has been processed by the business"* [4].

### 4.2 Can Consumers Successfully Exercise Their Opt-Out Rights?

Figure 2 outlines the process for evaluating opt-out request outcomes. First, we verified whether the request was successfully submitted. If it was not, it was marked as a "Failure to submit". For successfully submitted requests, we completed any required follow-up actions. If the request was later rejected or remained unresolved after two months, it was labeled as a "Failure to fulfill". It is important to note that some failures occurred because users could not provide the necessary personal information or were redirected to third-party advertising industry opt-out tools. These failures may not be considered CCPA violations.

| Status | Sub-status | Description | #Websites |
|---|---|---|---|
| Failure to submit | Malfunctioning opt-out page | Errors when rendering opt-out page or submitting request. | 13 |
| | Require offline opt-out steps | Website required users to email or call them to make a request. | 10 |
| | Instructions only | The opt-out link redirected users to a confusing page with vague instructions on completing the process. | 8 |
| | Requirement login | Users have to login to opt-out. | 4 |
| | Require unavailable personal information | Website required other information not available in the fake profile created. | 1 |
| | Advertising industry opt-out tool | The opt-out link redirected users to a time-consuming advertising industry opt-out tool that did not guarantee success. | 1 |
| Failure to fulfill | Ambiguous status | Websites either failed to notify or used vague wording, leaving the request status unknown. | 35 |
| | Outstanding follow-up actions | Websites required additional opt-out actions, like cookie settings or unavailable information from the fake profile. | 7 |
| | Incomplete within CCPA timeline | Request remained incomplete after 2 months of submission. | 6 |
| Resolved | Request confirmed to be successful | Website confirmed the request is successful. | 193 |
| | Non-matching profile | Website searched their database but could not find the profile. | 51 |
| | No action required | Website did not sell consumer personal information hence no action is required. | 1 |

Table 1. Outcomes of opt-out requests. In total opt-out requests had a "Failure to submit" on 37 websites (11.2%), "Failure to fulfill" for 48 websites (14.5%) and were resolved for 74.2%. N= 330.

We identified 37 instances where the opt-out request was not submitted. The main issues were that the opt-out page was either broken, provided only vague instructions, or redirected users to alternative methods such as using advertising industry opt-out tools, sending emails, or making phone calls. Additionally, we found 48 cases where the request was submitted but remained incomplete or had an unknown status. The primary issue was "Unclear status updates", where websites used vague language, leaving users unsure about the outcome of their request. This ambiguity may be a tactic employed by businesses to avoid fulfilling the request.

Of the 330 websites examined, only 245 (74.2%) successfully resolved the opt-out requests. Of these, 193 confirmed that the opt-out preferences were recorded, while 51 reported being unable to find a matching profile, likely due to the use of a fake profile during testing. The variation in how requests are handled raises concerns about how businesses manage opt-out requests. For instance, some websites reject requests if no matching profile is found, suggesting they perform a

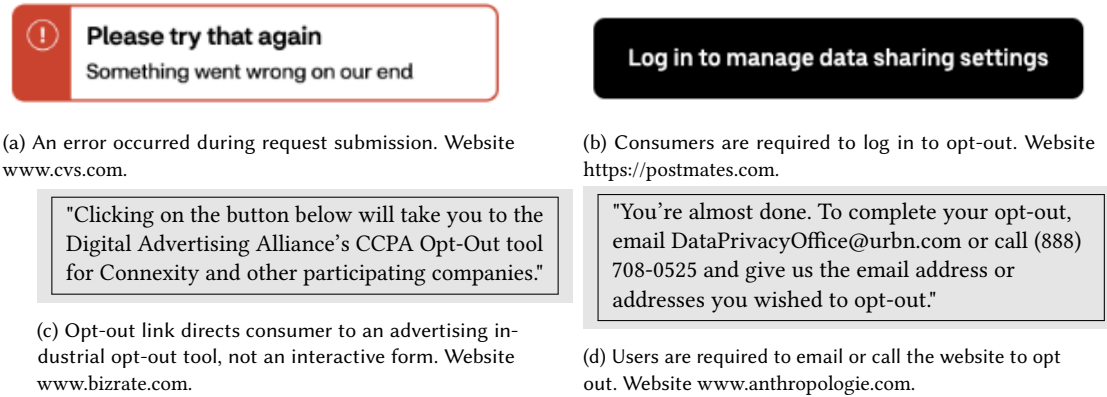

(a) An error occurred during request submission. Website www.cvs.com.

(b) Consumers are required to log in to opt-out. Website https://postmates.com.

"Clicking on the button below will take you to the Digital Advertising Alliance's CCPA Opt-Out tool for Connexity and other participating companies."

(c) Opt-out link directs consumer to an advertising industrial opt-out tool, not an interactive form. Website www.bizrate.com.

"You're almost done. To complete your opt-out, email DataPrivacyOffice@urbn.com or call (888) 708-0525 and give us the email address or addresses you wished to opt-out."

(d) Users are required to email or call the website to opt out. Website www.anthropologie.com.

Fig. 3. Examples of "Failure to submit" an opt-out request

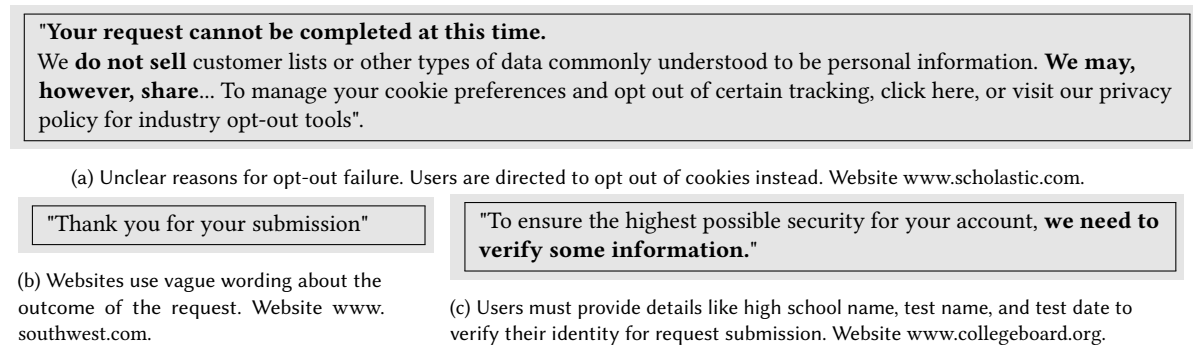

"**Your request cannot be completed at this time.**
We **do not sell** customer lists or other types of data commonly understood to be personal information. **We may, however, share**... To manage your cookie preferences and opt out of certain tracking, click here, or visit our privacy policy for industry opt-out tools".

(a) Unclear reasons for opt-out failure. Users are directed to opt out of cookies instead. Website www.scholastic.com.

"Thank you for your submission"

(b) Websites use vague wording about the outcome of the request. Website www.southwest.com.

"To ensure the highest possible security for your account, **we need to verify some information.**"

(c) Users must provide details like high school name, test name, and test date to verify their identity for request submission. Website www.collegeboard.org.

Fig. 4. Examples of "Failure to fulfill" an opt-out request.

database search in line with CCPA requirements. However, this raises questions about how requests from users without profiles are handled. Additionally, when websites confirm successful opt-out requests for non-existent profiles, it is unclear whether they genuinely conducted a search or merely pretended to comply with CCPA requirements. For more details on the description and frequency of opt-out failures, refer to Table 1. Figure 2 provides examples of "Failure to submit", and Figure 3 illustrates "Failure to fulfill".

We also analyzed the number of failures by website category (see Table 9 in the Appendix). Failures were found across nearly all categories, except for "ScienceMath". Some categories, such as "Electronics", "Apparel", and "HomeGarden", had higher failure rates, while others, like "Entertainment" and "Retail" had lower failure rates. This variation may be due to several factors. Categories with higher failure rates may rely more on data sales and sharing for targeted advertising and consumer analytics, making them less likely to fully comply with opt-out mechanisms. Conversely, categories with lower failure rates may focus more on direct customer engagement and rely less on third-party data sharing, leading them to prioritize compliance with opt-out mechanisms to build trust and maintain strong customer relationships.

| Opt-out requirements | Third-party | | | #Websites |
|---|---|---|---|---|
| | OneTrust | Others | None | |
| Opt-out form | 41 | 16 | 70 | 127 (43.5%) |
| Browser opt-out | 34 | 3 | 75 | 112 (38.4%) |
| Both control methods | 34 | 3 | 16 | 53 (18.1%) |
| Sum | 109 (37.3%) | 22 (7.5%) | 161 (55.1%) | 292 |

Table 2. Websites' opt-out control methods and third-party platforms supporting them. N = 292.

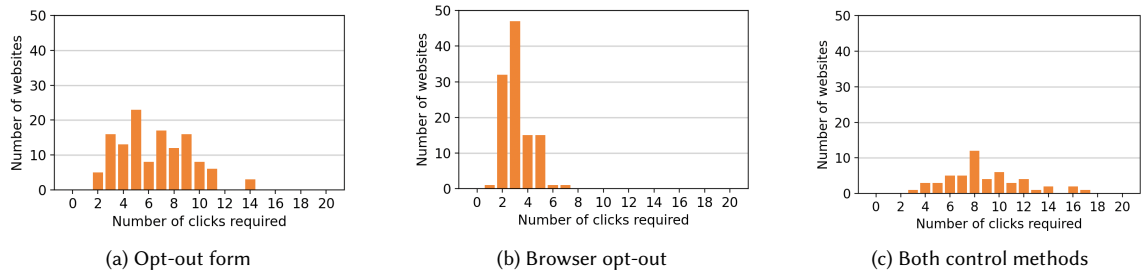

(a) Opt-out form  (b) Browser opt-out  (c) Both control methods

Fig. 5. Number of clicks required to submit an opt-out request for each type of opt-out control method. N=292.

### 4.3 What Does It Take for a Consumer to Submit an Opt-Out Request?

**1. Opt-out control methods:** We analyzed the opt-out implementation patterns of 292 websites where opt-out requests were successfully submitted. This analysis excludes 37 websites categorized as "Failure to Submit" and 1 website categorized as "Resolved" but marked as "No Action Required" (refer to Table 1 for details about these websites). The most common control method was the "Opt-out form" (43.5%), followed by "Browser opt-out" (38.4%), and "Both control methods" (18.2%). Some websites use third-party platforms to manage these opt-out processes, and we tracked which platforms were used. We found that 44.8% of the websites relied on third-party platforms, with "OneTrust" being the dominant service, accounting for 83.2% of those cases. Due to its significant presence, we treated "OneTrust" as a separate category. Websites using either the "Browser opt-out" or "Opt-out form" alone were less likely to use third-party services, while those employing "Both control methods" were more likely to rely on "OneTrust". For more details about opt-out control methods, refer to Table 2.

**2. Number of clicks:** Figure 5 shows the number of clicks required to submit an opt-out request for each type of control method. Websites using "Browser opt-out" generally require fewer clicks, with less variation, compared to those using "Opt-out form" or both methods. This suggests that "Browser opt-out" is not only less cumbersome but also more consistently implemented than the other methods.

**3. Personal information required:** Overall, 177 websites (60.6%) require consumers to provide personal information, although the amount of information required varies. Nearly all ask for an email address, likely for consumer identification. Additionally, 66 websites (37.3% of the 177 websites) request a physical address, and 44 websites (24.9% of the 177 websites) ask for a phone number. For more details on the types of personal information required, see Table 3.

|      | Name | Email | Address | Phone# | #Websites |
|------|------|-------|---------|--------|-----------|
|      | ○ | ○ | ○ | ○ | 115 (39.4%) |
|      | ○ | ● | ○ | ○ | 26 (8.9%) |
|      | ○ | ○ | ● | ○ | 1 (0.3%) |
|      | ● | ● | ○ | ○ | 76 (26.0%) |
|      | ○ | ● | ● | ○ | 1 (0.3%) |
|      | ● | ● | ● | ○ | 29 (9.9%) |
|      | ● | ● | ○ | ● | 9 (3.1%) |
|      | ● | ● | ● | ● | 34 (11.6%) |
| Sum  | 149 | 176 | 66 | 44 | 292 |

Table 3. Personal information required to submit an opt-out request. N=292.

### 4.4 Do Opt-Out Processes Violate CCPA Requirements?

We compared the reasons for opt-out failures with CCPA regulations to determine whether these failures might constitute CCPA violations (refer to Table 4 for a summary of the results). Our analysis revealed that five out of nine types of failures mentioned in Table 4 appear to breach explicit CCPA requirements. For example, the CCPA mandates that the opt-out process be *"functional"*, yet some websites feature broken opt-out pages. Additionally, some sites only provide users with vague instructions to opt-out or require them to complete offline steps, such as sending an email or making a phone call, instead of offering an *"interactive form"* for online submission. These offline methods complicate the opt-out process, as the steps are often unclear, leaving users unsure of the time and effort needed to complete their request.

In addition to the websites that appear to violate clear CCPA requirements, others may also be in violation, though this is harder to confirm due to the CCPA's vague and sometimes non-specific guidelines. For example, some websites request additional personal information (other than email, phone number, or address) during the opt-out process. While it is unclear if this violates the CCPA—since determining if this extra information is truly *"necessary"* is difficult—most websites do not require it. This suggests these requests may be intended to discourage opt-outs rather than serve legitimate purposes. Similarly, when websites require users to take extra steps, like adjusting privacy settings after submitting an opt-out request, it raises concerns about compliance. If the website already has the user's personal information, further actions seem unnecessary and may hint at misuse of these additional requirements.

Lastly, not notifying users or using vague language about the status of an opt-out request, while not a violation of CCPA requirements—since the law does not mandate businesses to confirm successful requests—can leave users uncertain about whether their opt-out was successful or if further action is needed. Without clear confirmation, it also becomes more difficult for users to hold businesses accountable for failing to honor their opt-out requests.

Regarding how opt-out process should be implemented, the CCPA requires businesses to offer an opt-out process that is *"easy to execute"* and involves a *"minimal number of steps"* [4]. However, since the CCPA does not specify which opt-out control methods meet these requirements, it is unclear whether a website's choice of control methods can be considered *"easy to execute"*. For instance, websites that require consumers to use both the opt-out form and browser opt-out raise questions about whether this dual action is necessary for fully opting out of the sale or sharing of personal information, or if it unnecessarily complicates the process.

| Status | Sub-status | CCPA non-compliance? | CCPA requirements violated |
|---|---|---|---|
| Failure to submit | Malfunctioning opt-out page | Y | Shall be "functional" |
| | Require offline opt-out steps | Y | Shall be an "interactive" form |
| | Instructions only | Y | Shall be an "interactive" form |
| | Require login | Y | Shall not require creating an account (to login) |
| | Require unavailable personal information | ? | Shall not require personal information beyond what is "necessary" to complete the request |
| | Advertising industry opt-out tool | ? | Not explicitly mentioned in the CCPA. |
| Failure to fulfill | Ambiguous status | N | CCPA makes it optional for businesses to confirm that the request is processed successfully |
| | Outstanding follow-up actions | ? | Require "minimal steps", shall not require personal information beyond what is "necessary" to complete the request |
| | Incomplete within CCPA timeline | Y | Comply with the request within 15 business days since the request is received |

Table 4. Opt-out requests' status and the CCPA requirements violated. For column "CCPA non-compliance?", "Y" indicates "clearly non-compliant", "N" indicates "compliant", "?" indicates "potentially non-compliant".

Similarly, since the law does not clearly define what constitutes a *"minimal number of steps"*, we set a threshold of 9 clicks (1 standard deviation above the average) to identify websites with overly tedious opt-out processes, which may violate the CCPA's *"easy to execute"* standard. This count excludes additional clicks for navigating fields, solving CAPTCHAs, or verifying identity via phone or email, indicating that this is the minimum number of clicks required to submit an opt-out request. Most websites using the "Browser opt-out" method stay within this threshold. Some using the "Opt-out form" exceed it while nearly half of the websites using "Both control methods" surpass this threshold.

The CCPA also mandates that businesses collect only the minimum personal information necessary to identify consumers. However, it is unclear whether requesting details like a physical address or phone number is truly *"necessary"* for completing the opt-out request or if these requirements are intentionally used to make the process more difficult and discourage opt-outs. This uncertainty raises questions about whether such practices violate CCPA regulations.

## 5 Findings About Dark Patterns

In this section, we present our findings on dark patterns in the opt-out process. First, we outline the CCPA's definition of dark patterns. Next, we describe the dark patterns identified in the opt-out process and assess whether they would

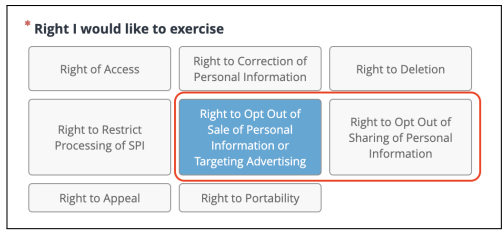
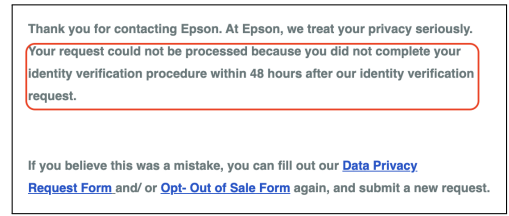

(a) *Separate Submissions* Users can only opt out of either the sale or sharing of their personal information per request—not both. Website www.finishline.com.

(b) *Identity Confirmation* Users must verify their identity within a timeframe. If they fail, they have to resubmit the request. Website https://epson.com/usa.

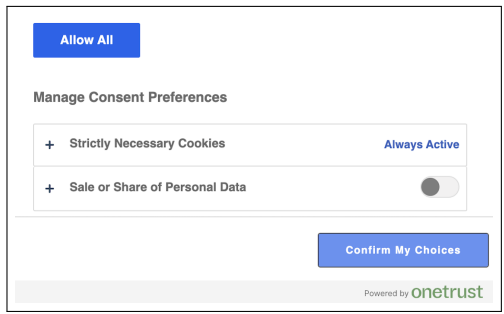
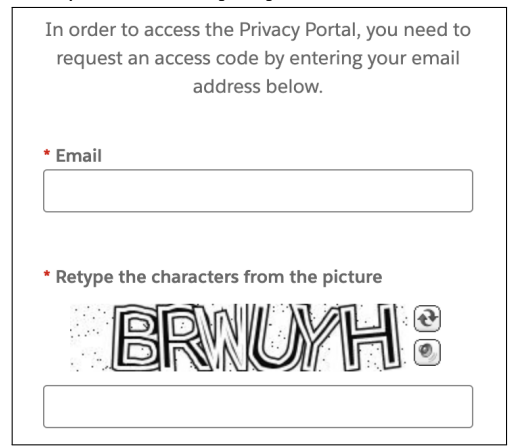

(c) *Asymmetry* It only takes one click for users to opt-in (by clicking "Allow All") but it takes 2 clicks to opt-out (toggle off and select "Confirm My Choices"). Website www.huffpost.com.

(d) *Verified email message* Users must solve a CAPTCHA and verify their identity via email to access a privacy portal. Website www.nfl.com.

Fig. 6. Examples of some "Obstruction" dark patterns.

qualify as dark patterns under the CCPA. Finally, we examine the characteristics of the websites that employ these dark patterns.

## 5.1 What Dark Patterns Are Prohibited by the CCPA?

The CCPA defines dark patterns as *"interfaces that substantially subvert or impair user autonomy, decision-making, and choice"* [4]. Additionally, the CCPA outlines five key principles, and any violation of these may be considered a dark pattern.

(1) *"Easy to understand. The methods shall use language that is easy for consumers to read and understand."*
(2) *"Symmetry in choice. The path for a consumer to exercise a more privacy-protective option shall not be longer or difficult or time-consuming than the path to exercise a less privacy-protective option."*
(3) *"Avoid language or interactive elements that are confusing to the consumer."*
(4) *"Avoid choice architecture that impairs or interferes with the consumer's ability to make a choice."*
(5) *"Easy to execute. The business shall not add unnecessary burden or friction to the process by which the consumer submits a CCPA request."*

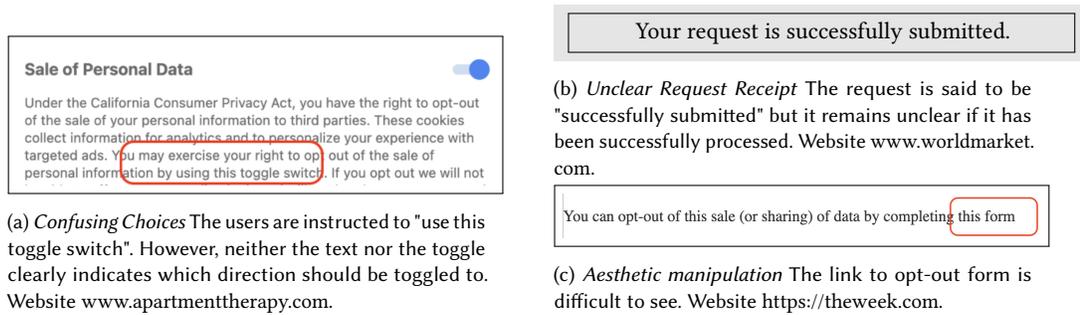

(a) *Confusing Choices* The users are instructed to "use this toggle switch". However, neither the text nor the toggle clearly indicates which direction should be toggled to. Website www.apartmenttherapy.com.

(b) *Unclear Request Receipt* The request is said to be "successfully submitted" but it remains unclear if it has been successfully processed. Website www.worldmarket.com.

(c) *Aesthetic manipulation* The link to opt-out form is difficult to see. Website https://theweek.com.

Fig. 7. Examples of some "Interface interference" and "Misdirection" dark patterns.

### 5.2 What Types of Dark Patterns Are Used in the Opt-Out Process?

Dark patterns in the opt-out process can be categorized into three high-level dark patterns: "Obstruction", "Interface interference" and "Misdirection".

**Obstruction:** "Obstruction" refers to dark patterns where users encounter barriers or hurdles that make it difficult for them to complete a task or access information [5]. Several types of "Obstruction" dark patterns were identified in the opt-out process, which we describe below.

- *Privacy maze:* This refers to an opt-out process that is unnecessarily complicated. We assess complexity by counting clicks, with nine or more—one standard deviation above the average—considered excessively complex.
- *Separate submissions:* This refers to websites that require users to submit separate requests for opting out of the sale and sharing of personal information, or for each service under a parent website, instead of offering a single, unified option. This makes the opt-out process unnecessarily complex and cumbersome.
- *CAPTCHA test:* Some websites make solving a CAPTCHA mandatory for consumers to submit an opt-out request. This adds an unnecessary burden to the opt-out process.
- *Asymmetry:* Some websites create obstruction by making it easier to opt into the sale or sharing of personal information than to opt out.
- *Identity confirmation:* Some websites require consumers to verify their identity via phone or email, either during or after submitting a request, often within a limited time window, before processing the request.
- *Verified email message:* Some websites communicate with consumers through a secure privacy portal instead of directly via email, making it tedious and difficult to read messages. For instance, to access a message sent through the privacy portal, a consumer must click on a link sent via email, solve a CAPTCHA challenge, and complete a time-sensitive email verification before they can open the message. Additionally, some websites require multiple messages for communication, making the follow-up process time-consuming and even more challenging.
- *Other actions:* After consumers submit an opt-out request, some websites may require them to perform additional actions that are more complex than simply clicking a verification link. These actions could include contacting a designated representative, providing additional personal information, or submitting another opt-out form.

**Interface interference:** Interface interference is a strategy used to manipulate interfaces, privileging certain actions and, thus, limiting the discoverability of alternatives [16]. Refer to Figure 7 for examples. This type of dark pattern includes the following:

- *Information overload:* Some websites require consumers to wade through a lengthy and confusing disclosure page which makes it time-consuming to locate the opt-out mechanisms.
- *Aesthetic manipulation:* Some opt-out mechanisms are difficult to locate due to their design choices. For example, some of the links to locate the opt-out mechanism might be too small, have bad color contrast with the surrounding text, and can be cluttered around the surrounding text.

**Misdirection:** This refers to dark patterns where the instructions about how to exercise privacy-protective choices are either not mentioned or are very confusing. Refer to Figure 7 for examples. They include the following:

- *Unclear request receipt:* For some websites, after a request is submitted, consumers receive either no notification or unclear instructions. As a result, they are often left uncertain about whether their opt-out request was successful or if additional steps are needed to complete the process. This lack of clarity can hinder consumers from fully completing the opt-out process, as they may be unaware of essential steps required to finalize their request.
- *Confusing choices:* The instructions for using the opt-out mechanism are unclear or confusing. This includes the use of double negatives, lack of guidance on how to operate the opt-out toggles, and vague descriptions of the opt-out form. This type of dark pattern is explicitly addressed in the CCPA.

### 5.3 Are These Dark Patterns Violating CCPA Requirements, or Are They Exploiting the Loopholes That Have Not Been Addressed by the CCPA?

We compared the dark patterns found in the opt-out process with CCPA regulations. Comparison details can be found in Table 5. We identified two patterns—"CAPTCHA test" and "Unclear request receipt"—that are not classified as dark patterns under the CCPA. "CAPTCHA test" is not mentioned in the CCPA, while "Unclear request receipt" is implicitly allowed, as businesses are not required to provide confirmation of the request's status. However, this lack of confirmation can leave consumers uncertain about whether further action is needed or if the request is complete. This pattern was found in 62 out of 292 websites (21.2%), raising concerns that businesses may be exploiting this ambiguity.

Five additional dark patterns we observe could potentially constitute dark patterns under the CCPA. However, due to the lack of clear definitions in the law and the subjective nature of these patterns, it is challenging to determine if they violate CCPA regulations. For instance, assessing the legal status of a "Privacy maze" dark pattern is difficult because the CCPA does not clearly define what constitutes *"minimal steps"*. Websites may argue that the steps they include are necessary for their opt-out process, making it hard to definitively label them as dark patterns. Similarly, the "Information overload" dark pattern is hard to assess, as what some users see as excessive information might seem reasonable to others.

We identified six dark patterns prohibited by the CCPA, with the most common being "Identity confirmation" (found on 71 websites) and "Verified email message" (found on 41 websites). Both involve follow-up actions after the request is submitted. Compared to "Verified email message," which requires users to authenticate themselves each time they access the privacy portal, "Identity confirmation" is less burdensome, typically requiring just clicking on a link sent via email. However, for both patterns, if users fail to authenticate within a short time frame, they may need to resubmit their request, adding to the burden. Despite being prohibited, these patterns remain prevalent, indicating that many websites are not complying with CCPA regulations. Additionally, because these patterns emerge during the follow-up process, they are harder to detect upfront, only becoming evident after a full opt-out request is submitted.

When analyzing how dark patterns violate the five CCPA-defined principles, we found that more than half of these patterns make the opt-out process less *"easy to execute"*. This indicates that many websites use tactics to complicate the process, likely to discourage consumers from opting out. We also found that several dark patterns, such as "Confusing choices" and "Information overload", infringe upon more than one CCPA principle. Interestingly, patterns such as the "CAPTCHA test" and "Unclear request receipt"—while not explicitly banned by the CCPA—still violate at least one principle. For example, "Unclear request receipt" may breach the requirement that the opt-out process be *"easy to understand"* and avoid manipulative language or design. This underscores potential loopholes in the CCPA, allowing businesses to exploit dark patterns that, while not directly forbidden, still hinder consumers' ability to opt out effectively.

| High level | Dark pattern | Easy to understand? | Symmetry in choice? | Avoid manipulative design? | Avoid manipulative language? | Easy to execute? | CCPA non-compliance? | Non-compliance section |
|---|---|---|---|---|---|---|---|---|
| Obstruction | Privacy maze | - | - | - | - | ✗ | ? | Require minimal steps |
| | Separate submissions | - | - | - | - | ✗ | Y | A single option to opt-out of the sale or sharing of all personal information |
| | CAPTCHA test | - | - | - | - | ✗ | N | Not explicitly mentioned in CCPA |
| | Asymmetry | - | ✗ | - | - | ✗ | Y | Symmetry in choice |
| | Identity confirmation | - | - | - | - | ✗ | Y | Shall not require a consumer to verify their identity |
| | Verified email message | - | - | - | - | ✗ | Y | Shall not require a consumer to verify their identity |
| | Other actions | - | - | - | - | ✗ | ? | Require minimal steps, information necessary to complete the request only |
| Interface interference | Information overload | ✗ | - | ✗ | - | - | ? | Impairs or interferes with the consumer's ability to make a choice |
| | Aesthetic manipulation | - | - | ✗ | - | - | ? | Impairs or interferes with the consumer's ability to make a choice. |
| Misdirection | Unclear request receipt | ✗ | - | - | ✗ | - | N | Not explicitly mentioned in CCPA |
| | Confusing choices | ✗ | - | ✗ | ✗ | - | ? | Toggles or buttons must clearly indicate the consumer's choice. |
| | Cookie control misdirection | ✗ | - | - | ✗ | - | Y | Notification or tool regarding cookies is not by itself an acceptable method for submitting requests to opt-out of sale/sharing |

Table 5. Dark patterns, CCPA-defined principles violated, whether the dark pattern is non-compliant to the CCPA, and the CCPA requirements violated. For column "CCPA non-compliance?", "Y" indicates "clearly non-compliant", "N" indicates "compliant", "?" indicates "potentially non-compliant".

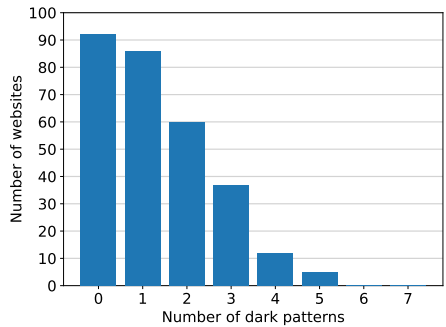

Fig. 8. Frequency of dark patterns across websites. N=292.

| High level | Dark pattern | #Websites |
|---|---|---|
| Obstruction | Identity confirmation | 71 (24.3%) |
|  | Privacy maze | 56 (19.2%) |
|  | Verified email message | 41 (14.0%) |
|  | Asymmetry | 24 (8.2%) |
|  | Separate submissions | 17 (5.8%) |
|  | CAPTCHA test | 8 (2.7%) |
|  | Other actions | 6 (2.1%) |
| Interface interference | Information overload | 41 (14.0%) |
|  | Aesthetic manipulation | 27 (9.2%) |
| Midirection | Unclear request receipt | 62 (21.2%) |
|  | Cookie control misdirection | 21 (9.2%) |
|  | Confusing choices | 19 (6.5%) |

Table 6. Prevalence of each dark pattern across websites. N=292.

## 5.4 What Are the Characteristics of the Websites That Employ Dark Patterns?

Figure 8 shows the distribution of dark patterns across websites. Our analysis shows that only 92 websites (31.5%) are completely free of dark patterns. In contrast, the majority of websites include at least one dark pattern, with the most common occurrence being a single dark pattern, followed by two dark patterns. Overall, fewer websites employ multiple dark patterns.

Overall, we found that 127 websites (43.5%) use at least one "Obstruction" dark pattern, 58 websites (19.9%) use at least one "Interface Interference" dark pattern, and 101 websites (34.6%) use at least one "Misdirection" dark pattern. For a detailed breakdown of each specific dark pattern, refer to Table 6. The frequent use of "Obstruction" dark patterns, combined with the range of tactics in this category, highlights how common these techniques are in the opt-out process. "Misdirection" dark patterns are also widely used, particularly the "Unclear request receipt" pattern. Although "Interface Interference" appears on many websites, it is less common compared to the other two categories.

We analyzed dark patterns across various opt-out control methods and third-party platforms to assess whether the use of dark patterns differs based on these factors. Our findings are summarized below.

**1. Dark patterns across different opt-out control methods:** We found that websites using "Browser opt-out" tend to have the fewest dark patterns, with most having none or only one. Websites using "Opt-out form" have slightly more dark patterns, while websites that use "Both control methods" have the highest number of dark patterns. This suggests that more complex opt-out methods are associated with a greater prevalence of dark patterns. Additionally, the distribution of dark patterns within each opt-out control method varies widely. For example, some websites using "Opt-out form" have no dark patterns at all while some websites also using "Opt-out form" have as many as five dark patterns. Refer to Figure 9 for the distribution of dark patterns across each type of opt-out control method.

We found that different opt-out control methods are associated with specific dark patterns. For instance, "Unclear request receipt" is common in both the "Opt-out form" and "Both methods" but not in "Browser opt-out." This is likely because "Browser opt-out" doesn't involve follow-up actions, minimizing ambiguity about request status. Similarly, dark patterns like "Identity confirmation" and "Verified email message" are common with other opt-out methods but not with "Browser opt-out," as it doesn't require identity verification. On the other hand, "Browser opt-out" is more prone to "Cookie control misdirection," where users are diverted into managing cookies instead of opting out of data sharing. Another frequent dark pattern in "Browser opt-out" is "Asymmetry," where opting in requires fewer steps than opting out.

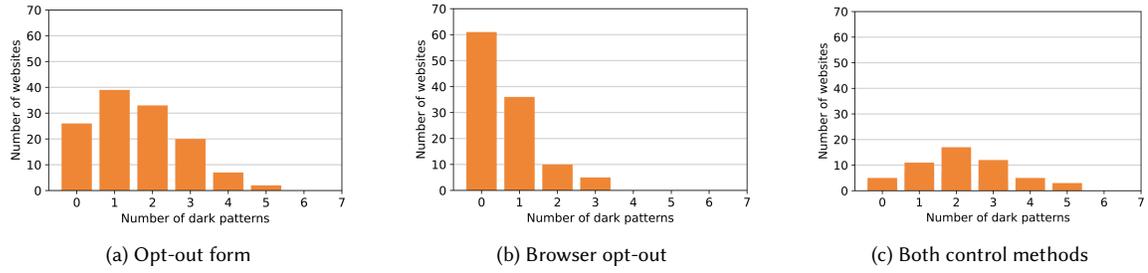

(a) Opt-out form　　　(b) Browser opt-out　　　(c) Both control methods

Fig. 9. Number of dark patterns found for each type of opt-out control method. N=292 websites.

| Ranking | Opt-out form | Browser opt-out | Both control methods |
| --- | --- | --- | --- |
| 1 | Identify confirmation (37.8%) | Cookie control misdirection (18.8%) | Privacy maze (43.4%) |
| 2 | Unclear request receipt (34.6%) | Asymmetry(14.3%) | Identity confirmation (37.7%) |
| 3 | Privacy maze (26.0%) | Information overload (13.4%) | Unclear request receipt (34.0%) |
| 4 | Verified email message (22.0%) | Confusing choices (12.5%) | Aesthetic manipulation (30.2%) |
| 5 | Information overload (13.4%) | Aesthetic manipulation (2.7%) | Verified email message (24.5%) |

Table 7. Top 5 dark patterns present in each type of opt-out control method.

**2. Dark patterns across different third-party platforms:** We found that websites not supported by third-party privacy platforms generally have fewer dark patterns compared to those that use third-party services like "OneTrust." This contradicts the expectation that privacy platforms, as compliance experts, would provide more standardized and CCPA-compliant solutions. Additionally, even among websites using the same platform, such as "OneTrust," the presence of dark patterns varies significantly. Some websites have none, while others have as many as 5 dark patterns. This suggests that third-party platforms may offer different versions of opt-out forms, potentially customized to fit the specific needs and design preferences of each website's opt-out process, although these design preferences might not necessarily comply with CCPA requirements. Refer to Figure 9 for more details about the prevalence of dark patterns across websites using different third-party platforms.

We also analyzed the most common dark patterns across different categories of third-party platforms. Some dark patterns, such as "Unclear request receipt," "Privacy maze," and "Identity confirmation," are prevalent across all third-party platforms, although their frequency varies by category. However, there are notable differences among the categories. For instance, "Cookie control misdirection" is common among websites that do not use third-party platforms but is rare among those that do. Additionally, "Verified email message" is present in 33% of websites using "OneTrust" but is uncommon in the other two categories. For more details on the top dark patterns across third-party platforms, refer to Table 8.

## 6 Discussion

In this section, we first discuss the implications of our study's findings. We then address the limitations and challenges in assessing CCPA compliance, both specific to this study and in general. Next, we provide specific recommendations for regulators, policymakers, and researchers in HCI, computer science, and related fields and suggest future directions that could benefit these disciplines.

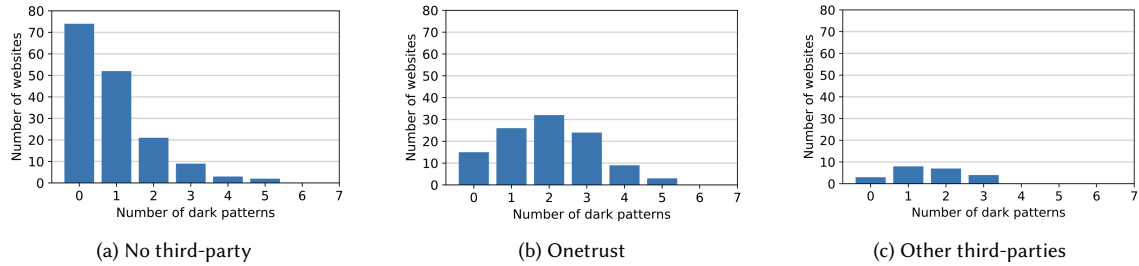

Fig. 10. Number of dark patterns found for each type of third-party platform. N=292.

| Ranking | No third-party | OneTrust | Other third-parties |
|---|---|---|---|
| 1 | Unclear request receipt (23.0%) | Identity confirmation (34.9%) | Identity confirmation (50.0%) |
| 2 | Information overload (13.7%) | Verified email message (33.0%) | Unclear request receipt (31.8%) |
| 3 | Identity confirmation (11.8%) | Privacy maze (29.4%) | Privacy maze (31.8%) |
| 4 | Privacy maze (10.6%) | Asymmetry (19.3%) | Information overload (18.2%) |
| 5 | Cookie control misdirection (9.9%) | Unclear request receipt (16.5%) | Aesthetic manipulation (13.6%) |

Table 8. Top 5 dark patterns present in each category of third-party platforms.

### 6.1 Effectiveness of the CCPA

We compared our study's findings with a 2021 study by Consumer Reports [3] on the CCPA opt-out process. Although the studies differ—Consumer Reports focused on the ability to execute opt-out rights, while we examined dark patterns in the opt-out process—our findings suggest that the opt-out process has improved since the CPRA came into effect. For example, Consumer Reports found that over 10% of data brokers required consumers to provide tedious information, such as government IDs, CA residency documents, selfies, or third-party apps, making the process burdensome. In contrast, we found very few websites requesting such personal information. Additionally, our study showed that 74% of opt-out requests were fully resolved, compared to only 40% in the Consumer Reports study. These findings suggest that CCPA compliance has improved since the implementation of the CPRA.

Despite the potential positive impacts of the CCPA on consumer privacy protection, our findings show that many websites still fail to comply with its regulations. In addition to websites lacking a functional opt-out form, some continue to use dark patterns explicitly prohibited by the CCPA. These issues highlight the need for stronger regulatory oversight and more effective enforcement to safeguard consumer privacy rights. Furthermore, the presence of dark patterns that are either unaddressed or implicitly accepted by the CCPA, as well as those with unclear compliance, suggests that websites may be exploiting ambiguities in the law to their advantage. This emphasizes the need for clearer definitions and careful evaluation of dark patterns to ensure robust consumer privacy protection.

## 6.2 Challenges of measuring CCPA non-compliance

Although stronger regulatory oversight is necessary to ensure CCPA compliance, enforcing it effectively poses several practical challenges. Below, we outline some of the difficulties we encountered during the study, which are likely shared by both regulators and HCI researchers.

*6.2.1 Human-Intensive and Time Consuming to Examine Opt-Out Process.* Our study shows that opt-out failures and dark patterns can occur at any stage of the process, requiring users to complete the entire opt-out procedure—submitting the request and following up on necessary actions—to determine if the request was fully resolved and identify dark patterns in the opt-out process. This is both time-consuming and labor-intensive. The challenge is compounded by the wide variation in how websites implement the opt-out process, requiring each site to be individually examined. This makes automated compliance checks difficult, further increasing the time and effort needed to assess CCPA compliance.

*6.2.2 Challenges of Defining a Dark Pattern.* Another challenge in measuring CCPA compliance is defining what qualifies as a dark pattern. Website developers may interpret CCPA requirements differently, potentially leading to the intentional or unintentional use of dark patterns that violates the regulations. This issue is further complicated by the vagueness of some CCPA guidelines, such as the requirement for businesses to implement an opt-out process with *"minimal steps"* and only request information *"necessary"* to identify consumers. Given that websites vary in the types and amounts of personal information collected, sold, or shared—and in how they handle opt-out requests—it becomes difficult to determine the appropriate opt-out controls, steps, and the amount of information to request from consumers. These factors make accurately evaluating compliance particularly challenging.

*6.2.3 Challenges of Verifying Whether Opt-Out Request Is Complied With and Fully Executed.* To assess CCPA compliance, it is essential to evaluate not only whether the opt-out process avoids dark patterns, but also whether it effectively stops the sale or sharing of personal information. This requires a clear understanding of the types of personal information collected, sold, or shared, who receives it, and how businesses prevent its sale or sharing. Even if a website's opt-out process meets CCPA requirements, the business would still be in violation if it continues to sell or share consumers' personal information without authorization. Verifying whether businesses genuinely stop the sale or sharing of data after an opt-out request is a complex task and an important area for future research. However, the lack of standardization in opt-out processes and the limited transparency in how businesses handle these requests make this task particularly challenging.

## 6.3 Recommendations

**1. For regulators:** The findings of this study provide clear evidence of CCPA violations, highlighting the need for regulators to take action and enforce the law. As discussed in Section 6.2, measuring compliance and enforcing the CCPA are complex and resource-intensive tasks. In the future, to investigate the opt-out processes of a larger number of websites, regulators might consider using approaches like crowdsourcing or automation. However, each approach has its limitations. Crowdsourcing can lead to inconsistencies and inaccuracies, while automation is rendered difficult by the wide variation in websites' opt-out processes.

**2. For policy makers:** The findings suggest that ambiguities and loopholes in the CCPA regulations may be exploited by apps or websites, particularly through dark patterns such as "Unclear request receipt" and "Privacy maze." These issues highlight the need for clearer guidelines and specific requirements to ensure that the law's objectives are not thwarted. For example, California's regulations prohibit user interface designs that thwart consumers' desires when they are making a choice to opt out. Yet the regulations do not have explicit language creating an obligation to confirm the consumer's choice to opt out after it has been made. As a result, consumers may spend several minutes opting out only to be left uncertain about whether their efforts succeeded. California's regulations prohibit all interactive elements that are confusing to consumers thus presumably rendering "unclear request receipt" dark patterns unlawful. However, the absence of an explicit requirement that consumers receive an unambiguous confirmation appears to create wiggle room that interface designers are exploiting, to the detriment of consumers. Additionally, we observed significant variation in the implementation of the opt-out process across websites. This variation makes automated compliance checking difficult, and without automation, enforcement becomes a tedious, labor-intensive process that is hard to scale. If not enforced effectively, the benefits of the regulations will be diminished. Therefore, it is crucial for policymakers to craft regulations that facilitate easier enforcement for regulators. Revisiting earlier regulations from time to time, after regulators learn what loopholes are being exploited, is a wise approach.

The findings also underscore the consequences of decisions that state legislators make when incorporating restrictions on dark patterns into state-level comprehensive privacy laws. California's restrictions on dark patterns appear to be moderately effective. Eighteen other states have enacted comprehensive consumer privacy laws [2], and by our count, six of them (Indiana, Iowa, Kentucky, Tennessee, Utah, and Virginia) lack specific provisions prohibiting the use of dark patterns in the opt-out process. Based on our observation that websites exploit loopholes even in a thoughtfully designed regulatory regime like California's, we expect dark patterns to regularly thwart user preferences to protect their privacy rights in jurisdictions where dark patterns remain largely unregulated. That said, because some websites choose not to provide different versions of their websites for residents of different states, as noted by Tran et al. [41], jurisdictions that fail to regulate dark patterns may to some extent free-ride on the efforts of California and other states.

Crafting regulations that are broad enough to prevent various types of dark patterns, yet specific enough to eliminate ambiguities and loopholes, is a challenging task. As the CCPA is the first law in the U.S. to explicitly prohibit the use of dark patterns, this challenge has been even more pronounced. Other jurisdictions tackling the dark patterns problem, both domestically and overseas, can learn from California's successes and failures. It is especially instructive to anticipate the opportunistic way in which some websites will try to comply with the letter, but not the spirit, of the CCPA's opt-out framework.

**3. For researchers:** Collaborations among legal experts and computer scientists can help society better understand both the interpretation of privacy laws and the data practices surrounding them. This paper illustrates how an interdisciplinary approach that leverages the strengths of different fields to tackle the complex and technically challenging task of privacy law compliance can contribute to both computer science and legal scholarship, while providing valuable guidance to policymakers and regulators.

## 7 Conclusion

The findings from this study indicate that while the CCPA opt-out process has improved since the CPRA took effect, there are still clear violations of CCPA requirements, and some websites may be exploiting ambiguities to implement dark patterns to their advantage. Dark patterns can appear not only during the submission of opt-out requests but also during the follow-up process after a request is submitted. This highlights the need for a thorough examination of the entire opt-out process to fully understand the extent of dark pattern use. The study also reveals significant variation in how websites implement the opt-out process, making compliance checking difficult for regulators. Therefore, policymakers should also focus on designing regulations that facilitate easier enforcement and anticipate circumvention strategies.

Yan Tran, Aarushi Mehrotra, Ranya Sharma, Marshini Chetty, Nick Feamster, Jens Frankenreiter, and Lior Strahilevitz

Dark Patterns in the Opt-Out Process and Compliance with the California Consumer Privacy Act (CCPA)				25[28] Eryn Ma and Eleanor Birrell. 2022. Prospective Consent: The Effect of Framing on Cookie Consent Decisions. In *Extended Abstracts of the 2022 CHI Conference on Human Factors in Computing Systems* (New Orleans, LA, USA) *(CHI EA '22)*. Association for Computing Machinery, New York, NY, USA, Article 400, 6 pages. https://doi.org/10.1145/3491101.3519687

[29] Dominique Machuletz and Rainer Böhme. 2020. Multiple Purposes, Multiple Problems: A User Study of Consent Dialogs after GDPR. *Proceedings on Privacy Enhancing Technologies* 2020, 2 (April 2020), 481–498. https://doi.org/10.2478/popets-2020-0037

[30] Arunesh Mathur, Gunes Acar, Michael J Friedman, Eli Lucherini, Jonathan Mayer, Marshini Chetty, and Arvind Narayanan. 2019. Dark patterns at scale: Findings from a crawl of 11K shopping websites. *Proceedings of the ACM on Human-Computer Interaction* 3, CSCW (2019), 1–32.

[31] Thomas Mildner, Gian-Luca Savino, Philip R Doyle, Benjamin R Cowan, and Rainer Malaka. 2023. About engaging and governing strategies: A thematic analysis of dark patterns in social networking services. In *Proceedings of the 2023 CHI conference on human factors in computing systems*. 1–15.

[32] Midas Nouwens, Ilaria Liccardi, Michael Veale, David Karger, and Lalana Kagal. 2020. Dark Patterns after the GDPR: Scraping Consent Pop-ups and Demonstrating their Influence. (04 2020).

[33] Midas Nouwens, Ilaria Liccardi, Michael Veale, David Karger, and Lalana Kagal. 2020. Dark patterns after the GDPR: Scraping consent pop-ups and demonstrating their influence. In *Proceedings of the 2020 CHI conference on human factors in computing systems*. 1–13.

[34] Sean O'Connor, Ryan Nurwono, Aden Siebel, and Eleanor Birrell. 2021. (Un) clear and (In) conspicuous: The right to opt-out of sale under CCPA. In *Proceedings of the 20th Workshop on Workshop on Privacy in the Electronic Society*. 59–72.

[35] Sean O'Connor, Ryan Nurwono, Aden Siebel, and Eleanor Birrell. 2021. (Un)clear and (In)conspicuous: The Right to Opt-out of Sale under CCPA. In *Proceedings of the 20th Workshop on Workshop on Privacy in the Electronic Society* (Virtual Event, Republic of Korea) *(WPES '21)*. Association for Computing Machinery, New York, NY, USA, 59–72. https://doi.org/10.1145/3463676.3485598

[36] Irwin Reyes, Primal Wijesekera, Abbas Razaghpanah, Joel Reardon, Narseo Vallina-Rodriguez, Serge Egelman, Christian Kreibich, et al. 2017. " Is Our Children's Apps Learning?" Automatically Detecting COPPA Violations. In *Workshop on Technology and Consumer Protection (ConPro 2017), in conjunction with the 38th IEEE Symposium on Security and Privacy (IEEE S&P 2017)*.

[37] Irwin Reyes, Primal Wijesekera, Joel Reardon, Amit Elazari Bar On, Abbas Razaghpanah, Narseo Vallina-Rodriguez, Serge Egelman, et al. 2018. "Won't somebody think of the children?" examining COPPA compliance at scale. In *The 18th Privacy Enhancing Technologies Symposium (PETS 2018)*.

[38] Iskander Sanchez-Rola, Matteo Dell'Amico, Platon Kotzias, Davide Balzarotti, Leyla Bilge, Pierre-Antoine Vervier, and Igor Santos. 2019. Can I opt out yet? GDPR and the global illusion of cookie control. In *Proceedings of the 2019 ACM Asia conference on computer and communications security*. 340–351.

[39] Aden Siebel and Eleanor Birrell. 2022. The Impact of Visibility on the Right to Opt-out of Sale under CCPA. *arXiv preprint arXiv:2206.10545* (2022).

[40] Than Htut Soe, Oda Elise Nordberg, Frode Guribye, and Marija Slavkovik. 2020. Circumvention by design - dark patterns in cookie consent for online news outlets. In *Proceedings of the 11th Nordic Conference on Human-Computer Interaction: Shaping Experiences, Shaping Society* (Tallinn, Estonia) *(NordiCHI '20)*. Association for Computing Machinery, New York, NY, USA, Article 19, 12 pages. https://doi.org/10.1145/3419249.3420132

[41] Van Hong Tran, Aarushi Mehrotra, Marshini Chetty, Nick Feamster, Jens Frankenreiter, and Lior Strahilevitz. 2024. Measuring Compliance with the California Consumer Privacy Act Over Space and Time. In *Proceedings of the CHI Conference on Human Factors in Computing Systems (CHI '24)*. ACM. https://doi.org/10.1145/3613904.3642597

[42] Christine Utz, Martin Degeling, Sascha Fahl, Florian Schaub, and Thorsten Holz. 2019. (Un)informed Consent: Studying GDPR Consent Notices in the Field. In *Proceedings of the 2019 ACM SIGSAC Conference on Computer and Communications Security* (London, United Kingdom) *(CCS '19)*. Association for Computing Machinery, New York, NY, USA, 973–990. https://doi.org/10.1145/3319535.3354212

[43] Maggie Van Nortwick and Christo Wilson. 2022. Setting the bar low: are websites complying with the minimum requirements of the CCPA? *Proceedings on Privacy Enhancing Technologies* (2022).

[44] Sebastian Zimmeck, Oliver Wang, Kuba Alicki, Jocelyn Wang, and Sophie Eng. 2023. Usability and enforceability of global privacy control. *Proceedings on Privacy Enhancing Technologies* 2 (2023), 1–17.
Manuscript submitted to ACM

# A APPENDIX

| Category | Failure to submit | Failure to fulfill | Total failure | #Websites | % Failure |
|---|---|---|---|---|---|
| Apparel | 5 | 11 | 16 | 43 | 37.2 |
| Finance | 2 | 2 | 4 | 27 | 14.8 |
| Automative | 2 | 2 | 4 | 26 | 15.4 |
| Retail | 0 | 2 | 2 | 25 | 8.0 |
| Electronics | 7 | 4 | 11 | 22 | 50.0 |
| Entertainment | 1 | 1 | 2 | 21 | 9.5 |
| Travel | 2 | 3 | 5 | 20 | 25.0 |
| FoodBeverage | 1 | 3 | 4 | 20 | 20.0 |
| News Media | 3 | 2 | 5 | 18 | 27.8 |
| HomeGarden | 3 | 4 | 7 | 17 | 41.2 |
| CareerEducation | 2 | 1 | 3 | 17 | 17.6 |
| HealthFitness | 3 | 0 | 3 | 14 | 21.4 |
| Beauty | 0 | 3 | 3 | 12 | 25.0 |
| Sports | 0 | 2 | 2 | 11 | 18.2 |
| Publishing | 1 | 3 | 4 | 9 | 44.4 |
| Reference | 3 | 1 | 4 | 8 | 50.0 |
| Telecommunication | 1 | 0 | 1 | 5 | 20.0 |
| PetsAnimals | 0 | 1 | 1 | 4 | 25.0 |
| ScienceMath | 0 | 0 | 0 | 4 | 0.0 |

Table 9. Number of failures for each category of websites, ordered by the number of websites in that category.

| High level | Dark pattern | $\kappa$ |
|---|---|---|
| Obstruction | Asymmetry | 0.93 |
| | Separate submissions | 0.54 |
| | Time-sensitive verification | 0.77 |
| | CAPTCHA test | 0.94 |
| Interface interference | Information overload | 0.36 |
| | Aesthetic manipulation | 0.41 |
| Midirection | Unclear request receipt | 0.72 |
| | Cookie control misdirection | 0.93 |
| | Confusing choices | 0.18 |

Table 10. Cohen's Kappa score for labels of dark patterns from Team Member 1 & 2.